\documentclass[conference]{IEEEtran}
\IEEEoverridecommandlockouts
\usepackage{cite}
\usepackage{amsmath,amssymb,amsfonts}
\usepackage{graphicx}
\usepackage{textcomp}
\usepackage{xcolor}
\usepackage{inconsolata}
\usepackage{listings}
\usepackage{booktabs}
\usepackage{subcaption}
\usepackage{url}
\DeclareUrlCommand\reg{\urlstyle{tt}}
\usepackage{cleveref}
\usepackage{tikz}
\usetikzlibrary{arrows.meta,positioning,calc}
\usepackage{pgfplots}
\pgfplotsset{compat=1.18}
\usepackage{multirow}

\def\BibTeX{{\rm B\kern-.05em{\sc i\kern-.025em b}\kern-.08em
    T\kern-.1667em\lower.7ex\hbox{E}\kern-.125emX}}

\begin{document}

\title{TLF: Rapid Characterization of RF Transceiver Parameters in Embedded Systems via Bus-Level Interception}

\author{
  \IEEEauthorblockN{Larry Hernandez}
  \IEEEauthorblockA{
    \textit{Dartmouth}\\
    Hanover, New Hampshire, USA\\
    l.gr@dartmouth.edu}
 \and
 \IEEEauthorblockN{Sergey Bratus}
 \IEEEauthorblockA{
   \textit{Dartmouth}\\
    Hanover, New Hampshire, USA\\
   Sergey.L.Bratus@dartmouth.edu}
 \thanks{Accepted for publication in the Proceedings of the 2026 IEEE
   Military Communications Conference (MILCOM 2026). This is the authors'
   accepted version; the final published version will appear in IEEE
   Xplore. \copyright~2026 IEEE. Personal use of this material is
   permitted. Permission from IEEE must be obtained for all other uses, in
   any current or future media, including reprinting/republishing this
   material for advertising or promotional purposes, creating new
   collective works, for resale or redistribution to servers or lists, or
   reuse of any copyrighted component of this work in other works.}
}

\maketitle

\bstctlcite{IEEEtranBSTCTLcontrol}

\begin{abstract}
We present \emph{TLF} (Transceiver Lifter Framework), a tool for
recovering RF transceiver configuration and runtime behavior from
bus-level traces captured between a microcontroller and its transceiver
IC\@.  A stateful protocol decoder, built against the transceiver's
register and data interface, reconstructs operating RF parameters and
behavior from intercepted register writes and FIFO transfers.  For
bus-attached transceivers whose hardware-cryptography keys are loaded
through the intercepted host interface, key material is also recoverable.
Where the
firmware drives frequency hopping---either through a hardware-assisted
engine or a custom schedule---the decoder extracts the channel table, hop
sequence, and timing.  We evaluate the approach on two targets from
different Semtech families: an SX1233-based UAV C2 modem employing
firmware-level FHSS with per-packet sync word rotation, and an
SX1276-based Meshtastic node exercising the LoRa register overlay.  From a
single bus capture, processed in seconds, TLF recovers the complete
register-exposed RF configuration (modulation, band plan, phase behavior)
without prior knowledge of the target firmware---sufficient to configure a
matched receiver or develop targeted countermeasures.  Above the chip
layer, a pluggable protocol decoder interprets recovered FIFO payloads as
application PDUs, demonstrated end-to-end on Meshtastic.
Firmware-level cryptographic state remains, as expected, opaque.  The
approach requires physical access or emulation of the target hardware,
and its recovery depth is bounded by the transceiver's register
interface: parameters implemented entirely in firmware (custom FEC,
whitening, encryption) are observable only as opaque FIFO payloads.
\end{abstract}

\begin{IEEEkeywords}
reverse engineering, RF transceivers, SPI, bus-level interception, FHSS,
embedded systems, counter-UAS, electronic warfare,
technical exploitation, LoRa, Meshtastic, protocol decoding
\end{IEEEkeywords}

\section{Introduction}\label{sec:introduction}

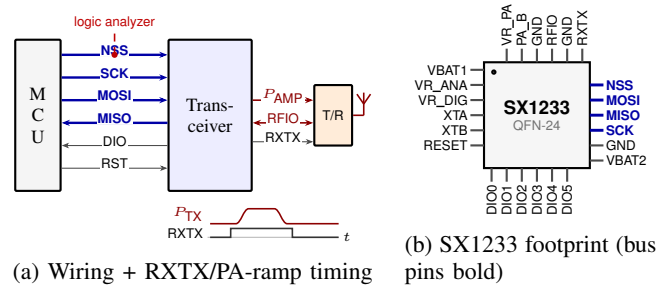
\begin{figure}[!t]
  \centering
  \begin{subfigure}[b]{0.60\columnwidth}
    \centering
    \begin{tikzpicture}[
      every node/.style={inner sep=1pt},
      ic/.style={draw, thick, align=center, rounded corners=0.5pt,
                 font=\scriptsize},
      spi/.style={-{Stealth[length=1.1mm]}, line width=0.9pt, blue!65!black},
      spirev/.style={{Stealth[length=1.1mm]}-, line width=0.9pt, blue!65!black},
      ctl/.style={-{Stealth[length=1mm]}, thin, black!70},
      ctlrev/.style={{Stealth[length=1mm]}-, thin, black!70},
      rf/.style={semithick, red!55!black},
      rfar/.style={-{Stealth[length=1.1mm]}, semithick, red!55!black},
      spilbl/.style={font=\tiny\sffamily\bfseries, blue!65!black,
                     fill=white, inner sep=0.3pt},
      pinlbl/.style={font=\tiny\sffamily, fill=white, inner sep=0.3pt},
      rflbl/.style={font=\tiny\sffamily, text=red!55!black,
                    fill=white, inner sep=0.3pt},
    ]
      \node[ic, fill=gray!8, minimum width=6mm, minimum height=20mm] (mcu)
        {M\\C\\U};
      \node[ic, fill=blue!7, minimum width=11mm, minimum height=20mm,
            right=14mm of mcu] (trx) {Trans-\\ceiver};
      \node[ic, fill=orange!15, minimum width=5mm, minimum height=8mm,
            right=8mm of trx, font=\tiny\sffamily, align=center] (sw)
        {T/R};

      \foreach \sig/\y/\arr in {NSS/8/spi,SCK/5/spi,MOSI/2/spi,MISO/{-1}/spirev}{%
        \draw[\arr] ([yshift=\y mm]mcu.east) -- ([yshift=\y mm]trx.west);
        \node[spilbl]
          at ($(mcu.east)!0.5!(trx.west)+(0,\y mm + 0.6mm)$) {\sig};
      }
      \draw[ctlrev] ([yshift=-4mm]mcu.east) -- ([yshift=-4mm]trx.west);
      \node[pinlbl] at ($(mcu.east)!0.5!(trx.west)+(0,-3.4mm)$) {DIO};
      \draw[ctl] ([yshift=-7mm]mcu.east) -- ([yshift=-7mm]trx.west);
      \node[pinlbl] at ($(mcu.east)!0.5!(trx.west)+(0,-6.4mm)$) {RST};

      \draw[rfar] ([yshift=2mm]trx.east) -- ([yshift=2mm]sw.west);
      \node[rflbl] at ($(trx.east)!0.5!(sw.west)+(0,2.6mm)$) {$P_\text{AMP}$};
      \draw[rf, {Stealth[length=1.1mm]}-{Stealth[length=1.1mm]}]
        ([yshift=-1mm]trx.east) -- ([yshift=-1mm]sw.west);
      \node[rflbl] at ($(trx.east)!0.5!(sw.west)+(0,-0.4mm)$) {RFIO};
      \draw[ctl] ([yshift=-3.5mm]trx.east) -- ([yshift=-3.5mm]sw.west);
      \node[pinlbl] at ($(trx.east)!0.5!(sw.west)+(0,-2.9mm)$) {RXTX};

      \draw[rf] (sw.east) -- ++(1.4mm,0) coordinate (a);
      \draw[rf] (a) -- ++(0,1.5mm) coordinate (atop);
      \draw[rf] (atop) -- ++(-0.9mm,1.3mm);
      \draw[rf] (atop) -- ++(0.9mm,1.3mm);
      \draw[rf] (atop) -- ++(0,1.5mm);

      \coordinate (tap) at ($(mcu.east)!0.5!(trx.west)+(0,8mm)$);
      \fill[red!75!black] (tap) circle (0.4mm);
      \draw[red!75!black, semithick, densely dashed]
        (tap) -- ++(0,3mm)
        node[above, font=\tiny\sffamily, red!75!black] {logic analyzer};

      \begin{scope}[shift={($(trx.south)+(0,-6mm)$)}, x=1.7mm, y=0.9mm]
        \draw[thin, gray!55] (-0.2,0) -- (10,0);
        \node[font=\tiny\sffamily, anchor=east, inner sep=1pt]
          at (-0.3, 0.6) {RXTX};
        \draw[semithick, black!80]
          (0,0) -- (1.6,0) -- (1.6,1.3) -- (6.4,1.3) -- (6.4,0) -- (10,0);
        \node[font=\tiny\sffamily, anchor=east, inner sep=1pt,
              text=red!55!black] at (-0.3, 3.4) {$P_\text{TX}$};
        \draw[semithick, red!55!black]
          (0,2) -- (2,2)
          .. controls (2.4,2) and (2.6,4.2) .. (3,4.2)
          -- (5,4.2)
          .. controls (5.4,4.2) and (5.6,2) .. (6,2)
          -- (10,2);
        \node[font=\tiny\sffamily, anchor=west, inner sep=1pt]
          at (10.1, 0) {$t$};
      \end{scope}
    \end{tikzpicture}
    \caption{Wiring + RXTX/PA-ramp timing}
    \label{fig:mcu-trx-wiring}
  \end{subfigure}\hfill
  \begin{subfigure}[b]{0.38\columnwidth}
    \centering
    \begin{tikzpicture}[
      pin/.style={thick, black!70},
      bus/.style={line width=1.1pt, blue!65!black},
      pinL/.style={font=\tiny\sffamily, anchor=east, inner sep=0.8pt},
      pinR/.style={font=\tiny\sffamily, anchor=west, inner sep=0.8pt},
      pinT/.style={font=\tiny\sffamily, rotate=90,
                   anchor=west, inner sep=0.8pt},
      pinB/.style={font=\tiny\sffamily, rotate=90,
                   anchor=east, inner sep=0.8pt},
      busR/.style={font=\tiny\sffamily\bfseries, anchor=west,
                   text=blue!65!black, inner sep=0.8pt},
    ]
      \def\sz{1.4}
      \def\pl{0.18}
      \pgfmathsetmacro{\spc}{\sz/7}

      \draw[thick, fill=gray!7] (0,0) rectangle (\sz,\sz);
      \draw[thick] (0.13, \sz-0.13) circle (0.6pt);
      \node[font=\scriptsize\sffamily\bfseries]
        at (\sz/2, \sz/2+0.10) {SX1233};
      \node[font=\tiny\sffamily, text=black!55]
        at (\sz/2, \sz/2-0.13) {QFN-24};

      \foreach \i/\n in {1/{VBAT1}, 2/{VR\_ANA}, 3/{VR\_DIG},
                         4/{XTA}, 5/{XTB}, 6/{RESET}} {
        \pgfmathsetmacro{\y}{\sz - \i*\spc + \spc/2}
        \draw[pin] (0,\y) -- (-\pl,\y);
        \node[pinL] at (-\pl,\y) {\n};
      }
      \foreach \i/\n in {1/{VBAT2}, 2/{GND}} {
        \pgfmathsetmacro{\y}{\i*\spc - \spc/2}
        \draw[pin] (\sz,\y) -- (\sz+\pl,\y);
        \node[pinR] at (\sz+\pl,\y) {\n};
      }
      \foreach \i/\n in {3/{SCK}, 4/{MISO}, 5/{MOSI}, 6/{NSS}} {
        \pgfmathsetmacro{\y}{\i*\spc - \spc/2}
        \draw[bus] (\sz,\y) -- (\sz+\pl,\y);
        \node[busR] at (\sz+\pl,\y) {\n};
      }
      \foreach \i/\n in {1/{DIO0}, 2/{DIO1}, 3/{DIO2},
                         4/{DIO3}, 5/{DIO4}, 6/{DIO5}} {
        \pgfmathsetmacro{\x}{\i*\spc - \spc/2}
        \draw[pin] (\x,0) -- (\x,-\pl);
        \node[pinB] at (\x,-\pl) {\n};
      }
      \foreach \i/\n in {1/{RXTX}, 2/{GND}, 3/{RFIO},
                         4/{GND}, 5/{PA\_B}, 6/{VR\_PA}} {
        \pgfmathsetmacro{\x}{\sz - \i*\spc + \spc/2}
        \draw[pin] (\x,\sz) -- (\x,\sz+\pl);
        \node[pinT] at (\x,\sz+\pl) {\n};
      }
    \end{tikzpicture}
    \caption{SX1233 footprint (bus pins bold)}
    \label{fig:sx1233-footprint}
  \end{subfigure}
  \caption{(a) Canonical MCU--transceiver topology in RF-enabled
           embedded systems.  The SPI bus (bold) carries all
           configuration and FIFO traffic and is the bus-level
           interception target.  (b) SX1233 QFN-24 pinout (top view)
           with bus pins (\textbf{NSS}, \textbf{MOSI}, \textbf{MISO},
           \textbf{SCK}) emphasized.}
  \label{fig:mcu-trx-overview}
\end{figure}

Firmware reverse engineering~\cite{eilam2011reversing} is a slow, expert
discipline, particularly on embedded systems built around proprietary or
under-supported MCU architectures (e.g., AVR variants), which often
require specialized tooling~\cite{ea5d0946695447a98aa9fa3025c8f9ef}.  Even
once firmware is extracted and disassembled, understanding its logic
remains costly.  Signal-level approaches are equally challenging: SDR
capture and blind demodulation demand proximity, tolerate noise or interference
poorly, and---for FHSS systems---must solve hop tracking before any
single channel can be decoded.

These costs matter operationally for radio-frequency platforms,
particularly UAV and IoT command-and-control (C2) modems.  Recovered or
captured systems---including drones for
tactical use, whose RF stacks are undocumented and whose firmware is
typically read-out
protected~\cite{DIU2024SX1276LastochkaM,DIU2025LLCC68BM35,DIU2026KlynLoRaModem}---create
a recurring need: extract the RF
parameters of an adversary link quickly enough to deploy matched
receivers, targeted jamming~\cite{Wu2026PreambleInjectionJamming}, etc.  Shortening the path to operational understanding of the RF layer
also enables a high-fidelity sample corpus that can accelerate downstream
firmware RE.  Transceivers of exactly this class are often recovered from
fielded combat UAVs~\cite{DIU2024SX1276LastochkaM,DIU2026KlynLoRaModem},
while the same LoRa stacks are the subject of conflict-area communications
training and deployment
guidance~\cite{KrukUAVTrainingCenter,Suantak2023MyanmarMeshtastic}.

Provided a logic analyzer can be attached, intercepting MCU--transceiver
traffic over SPI rapidly exposes the operational RF configuration
directly---modulation, sync word, frequency plan, and any
hardware-cryptography keys loaded across it---without firmware-level
understanding.

\section{Our Contributions}\label{sec:contributions}

We make the following contributions in the form of TLF, a Python
framework for automated transceiver characterization from bus traces:

\begin{enumerate}
  \item \textbf{A stateful decoder architecture} for MCU--transceiver bus
    traces that maintains a shadow register file, classifies every SPI
    transaction (register read/write, FIFO burst, mode transition), and
    emits a time-stamped event stream (\cref{sec:decoders}).

  \item \textbf{A transceiver-agnostic intermediate model} for recovered
    RF state---modulation, frequency plan, sync word, preamble, FIFO
    payloads, encryption key usage---that supports diffing, serialization,
    and downstream consumption by tools such as GNU~Radio
    (\cref{sec:parameter-modeling}).

  \item \textbf{A pluggable application-layer decoding stage} that consumes
    chip-level recovered packets and parses them as protocol PDUs.  It is
    chip-agnostic, so RF recovery and protocol parsing are composable (\cref{sec:protocol-layer}).

  \item \textbf{A case-study evaluation} on an SX1233-based~\cite{semtech-sx1233} modem with
    firmware-driven FHSS and per-packet sync word rotation,
    demonstrating recovery of the register-exposed physical-layer
    configuration, beaconing protocol, site-survey behavior, and
    channel hop plan from a single bus capture (\cref{sec:cs1}); and a
    cross-family evaluation on an SX1276-based~\cite{semtech-sx1276}
    Meshtastic node exercising the LoRa overlay and the protocol layer
    end-to-end (\cref{sec:cs2}).

  \item \textbf{Enhancements to sigrok} enabling continuous capture
    streaming on resource-constrained, low-cost logic analyzers,
    contributed as an open-source patch~\cite{sigrok}.
\end{enumerate}

Together, these enable rapid, low-cost RF characterization independent of
target MCU architecture or firmware comprehension. Once a transceiver
family is supported, the effort and cost reduction amortizes across every future
target sharing that family.

\section{How Transceivers Operate in Embedded Systems}
\label{sec:transceiver-operation}

The predominant design pattern for RF-enabled embedded systems involves a
microcontroller (MCU) communicating with a dedicated RF transceiver over a bus such as SPI
(\cref{fig:mcu-trx-overview}).
Interrupt (IRQ) driven events and direct memory access (DMA) may also be involved, but the core
communication and control flow is typically centered around the MCU writing to and reading from
the transceiver's registers and FIFOs.

RF-enabled embedded systems split into two classes: (i) those that use the
transceiver's built-in modulation, encryption, and FHSS as a turnkey data
pipe (typical of low-cost IoT, e.g., ``textbook'' LoRa modems); and (ii)
those that implement custom behavior in firmware---FEC, whitening, FHSS
schedule, encryption above the transceiver layer---using the transceiver
purely as an RF front-end. Class~(ii) is typical of UAV C2 and other tactical
platforms~\cite{Ghazali2021UAVLoRaReview,Davoli2021HybridLoRaUAVSwarming,Cariou2022BuriedSensorUAV}, especially
those using COTS transceivers.

In class~(i), RF-level properties map directly to register
values and are fully recoverable from bus traffic. In class~(ii), recovery
covers everything the transceiver itself implements while firmware-level
behavior remains observable as opaque FIFO payloads---still useful as
ground truth for downstream firmware analysis.

\subsection{Configuration and Parameter Setting}
\label{sec:config-params}

After power-on reset the MCU writes an initialization sequence that
programs modulation, carrier frequency, bit rate, deviation, receiver
bandwidth, preamble, sync word, and packet format.  \Cref{tab:key-registers}
(\cref{sec:register-interception}) lists the registers most relevant to a
decoder for Semtech transceivers.  Monolithic modems may expose UART
externally, but communicate with the transceiver over SPI internally.

\subsection{Encryption}\label{sec:encryption}

Some transceiver ICs include a hardware cryptographic engine that
encrypts and decrypts payloads transparently between the FIFO and the
radio front-end.  \Cref{tab:aes-transceivers} surveys representative parts.
For parts in this class that expose on-chip key loading through
host-visible registers, the encryption key is written via the same SPI
interface used for all other configuration, which means a bus-level
capture that includes \emph{the key-write transaction reveals the key itself}.

\begin{table}[t]
  \centering
  \caption{Hardware encryption support in common RF transceivers.}
  \label{tab:aes-transceivers}
  \footnotesize
  \setlength{\tabcolsep}{3pt}
  \begin{tabular}{@{}llrlr@{}}
    \toprule
    \textbf{Part}    & \textbf{Vendor} & \textbf{Band (MHz)} & \textbf{AES} & \textbf{Key Interface} \\
    \midrule
    SX1233              & Semtech   & 433/868/915     & 128     & \texttt{0x3E--4D} \\
    SX1276/7/8/9$^\ast$ & Semtech   & 137--1020       & ---     & --- \\
    SX1261/2            & Semtech   & 150--960        & ---     & --- \\
    CC1101              & TI        & 300--928        & ---     & --- \\
    CC1200              & TI        & 136--960        & 128     & ext.\ \texttt{0x2FE0} \\
    nRF24L01+           & Nordic    & 2\,400          & ---     & --- \\
    nRF52840$^\dagger$  & Nordic    & 2\,400          & 128     & periph.\ ECB/CCM \\
    ADF7023$^\ddagger$  & ADI       & 431--928        & 128/256 & pkt.\ RAM \texttt{0x06A} \\
    Si4463              & SiLabs    & 142--1050       & ---     & --- \\
    A7139               & AMICCOM   & 100--950        & ---     & --- \\
    \bottomrule
    \multicolumn{5}{@{}p{0.97\columnwidth}@{}}{\scriptsize
      $^\ast$\,No AES engine in either the FSK/OOK or the LoRa modem.
      \texttt{0x3D--0x4D} hold low-battery, IRQ-flag, DIO-mapping,
      version, PLL-hop and TCXO registers.
      $^\dagger$\,SoC, not a bus-attached transceiver: AES is an on-die
      peripheral, so keys never cross an external bus.
      $^\ddagger$\,Key staged into packet RAM via a command interface, not
      a directly addressed bank.} \\
  \end{tabular}
\end{table}

On parts that expose such an engine through a directly addressed register
bank the key is loaded as ordinary SPI register writes---on the SX1233,
16 consecutive bytes to \reg{RegAesKey1--16}
(\texttt{0x3E}--\texttt{0x4D}).  These registers are \emph{write-only}:
the firmware cannot read the key back, which protects it from
software-level extraction.  However, this protection is ineffective
against bus-level interception, because the key bytes travel in the clear
on the MOSI line during the write transaction.  The key load typically
occurs once during initialization and is therefore captured in any trace
covering power-on. Runtime key rotation appears as subsequent key-write transactions.  The
\emph{AesOn} bit (\reg{RegPacketConfig2[0]}, \texttt{0x3D}) confirms the
engine is active.

Hardware AES is, however, rarer than the LoRa-centric literature suggests,
and the distinction matters for scoping an analysis.  The SX1276/77/78/79
family---widely fielded in both COTS RC links~\cite{ExpressLRSRepository,RadioMasterBandit,BetaFPVSuperP}
and recovered combat UAVs~\cite{DIU2024SX1276LastochkaM,DIU2026KlynLoRaModem,AiThinkerRa01H}---has
\emph{no} AES engine in either its FSK/OOK or its LoRa modem.  The address range
sometimes assumed to hold a key bank in fact holds low-battery, IRQ-flag,
DIO-mapping, version, PLL-hop and TCXO registers.  On SX127x targets
(\cref{sec:cs2}) any confidentiality is therefore implemented in firmware,
above the transceiver boundary, and its keys never reach the bus.

\subsection{FHSS and Spectrum Behavior}\label{sec:fhss}

On the supported SX1233 and SX127x mechanisms, each host-directed
frequency change is observable on the bus as a sequence of carrier-frequency
register writes, whether the hop is driven by a built-in engine or by
firmware.  The detailed mechanics of detection are
discussed in \cref{sec:sync-freq-detection}.

\section{Methodology}\label{sec:methodology}

Our workflow has four stages: (1) identify the MCU--transceiver bus by
inspecting the PCB and correlating candidate nets with datasheets and
reference designs (typically SPI); (2) instrument the bus with a logic
analyzer at a sampling rate that resolves individual transactions, ideally
capturing the initialization sequence so that configuration writes,
key-loading events, and early mode transitions are preserved;
(3) decode the recorded traffic with standard SPI
tooling~\cite{saleae,sigrok}, yielding a time-ordered stream of register
reads/writes, FIFO accesses, and mode changes; (4) apply a family-specific
decoder that maps these transactions onto transceiver semantics and
reconstructs RF-domain state and runtime behavior.

\section{Designing a Transceiver Protocol Decoder}
\label{sec:decoders}

A stateful decoder must model the transceiver's register semantics, mode
transitions, and FIFO protocol.  We build each decoder directly from the
manufacturer's datasheet~\cite{semtech-sx1233,semtech-sx1276}. Reverse
engineering the command interface is unnecessary because most parts---including
clones widely available under different names---replicate the register
map of known Semtech designs.

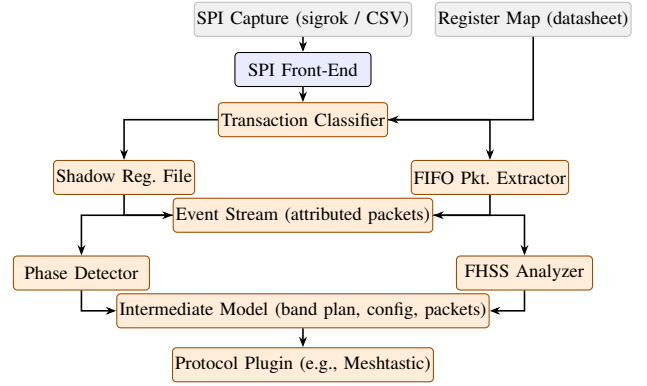
\begin{figure}[t]
  \centering
  \begin{tikzpicture}[
    node distance=2mm and 3mm,
    every node/.style={font=\scriptsize},
    block/.style={draw, rounded corners=1.5pt, minimum height=4.5mm,
                  minimum width=18mm, align=center, fill=blue!8,
                  inner sep=1pt},
    novel/.style={block, fill=orange!15, draw=orange!60!black},
    input/.style={block, fill=gray!12, draw=gray!50},
    arr/.style={-{Stealth[length=1.4mm]}, semithick},
  ]
    \node[input] (capture) {SPI Capture (sigrok / CSV)};
    \node[input, right=of capture] (datasheet) {Register Map (datasheet)};
    \node[block, below=of capture] (frontend) {SPI Front-End};
    \node[novel, below=of frontend] (classifier) {Transaction Classifier};
    \node[novel, below left=3mm and 3mm of classifier] (shadow)
      {Shadow Reg.\ File};
    \node[novel, below right=3mm and 3mm of classifier] (fifo)
      {FIFO Pkt.\ Extractor};
    \node[novel, below=8mm of classifier] (events)
      {Event Stream (attributed packets)};
    \node[novel, below left=3mm and 3mm of events] (phase)
      {Phase Detector};
    \node[novel, below right=3mm and 3mm of events] (fhss)
      {FHSS Analyzer};
    \node[novel, below=8mm of events] (model)
      {Intermediate Model (band plan, config, packets)};
    \node[novel, below=2.5mm of model] (proto)
      {Protocol Plugin (e.g., Meshtastic)};

    \draw[arr] (model) -- (proto);
    \draw[arr] (capture) -- (frontend);
    \draw[arr] (datasheet) |- (classifier);
    \draw[arr] (frontend) -- (classifier);
    \draw[arr] (classifier) -| (shadow);
    \draw[arr] (classifier) -| (fifo);
    \draw[arr] (shadow) |- (events);
    \draw[arr] (fifo) |- (events);
    \draw[arr] (events) -| (phase);
    \draw[arr] (events) -| (fhss);
    \draw[arr] (phase) |- (model);
    \draw[arr] (fhss) |- (model);
  \end{tikzpicture}
  \caption{TLF processing pipeline.  Gray blocks use standard tooling;
           orange blocks are novel contributions.  The shadow register file
           is the central data structure, enabling every FIFO or payload transaction to
           be attributed with the RF configuration active at that instant.}
  \label{fig:architecture}
\end{figure}

\Cref{fig:architecture} shows the TLF processing pipeline.  Raw SPI
captures (sigrok or CSV) are ingested by a front-end that produces a
time-ordered stream of SPI frames.  A transaction classifier separates
register accesses from FIFO bursts using the address byte's wnr bit and
the target address.  Register writes feed a shadow register file that
tracks the transceiver's full configuration at every point in time.  FIFO
transactions are paired with the active shadow state to produce
attributed packets (direction, frequency, sync word, payload).  Finally,
a phase detector and FHSS analyzer consume the event stream to identify
operating phases (beaconing, site survey, operational FHSS) and extract
band plans, hop timing, and sync word rotation statistics.  The shaded
blocks in \cref{fig:architecture} represent the novel components.  The SPI
front-end reuses standard tooling.

\subsection{Technical Challenges}\label{sec:challenges}

Register maps are public, but extracting operational RF behavior from
millions of raw SPI transactions required solving several non-trivial
problems:

\paragraph{FIFO vs.\ register disambiguation}
Both FIFO data and register~\texttt{0x00} share the same SPI address on
the SX1233 and SX1276.  A multi-byte write to address~\texttt{0x00} could
be a FIFO burst or a single register write followed by an auto-incremented
burst into~\texttt{0x01+}.  TLF resolves this by tracking the current data
mode: in packet mode, multi-byte accesses at~\texttt{0x00} are FIFO
operations; in continuous mode they are not.  The bit sits in different
places---\emph{DataMode} is in \reg{RegDataModul} (\texttt{0x02}) on the
SX1233 but \reg{RegPacketConfig2} (\texttt{0x31}) on the SX127x---so the
decoder consults whichever register its family map designates.  Burst
length and the preceding \reg{RegOpMode} write give further cues.  In LoRa
mode the ambiguity takes a different form: the modem exposes a 256-byte
data buffer through an address pointer, and in the captures evaluated here
each packet occupied a single qualifying burst.  That is target behavior,
not a hardware guarantee---in general a decoder must associate accesses by
address pointer, payload-length registers, IRQ events and mode transitions.

\paragraph{Packet boundary recovery}
The SPI bus carries interleaved register writes and FIFO bursts with no
explicit packet delimiters.  TLF reconstructs packet boundaries by
detecting the mode-transition pattern: a FIFO write burst bracketed by a
Standby$\to$Tx mode change constitutes one TX packet; a FIFO read burst
following an Rx$\to$Standby transition constitutes one RX packet.  This
heuristic correctly segmented all 682~packets in the case study without
manual intervention.

\paragraph{Automatic phase detection}
A raw capture contains distinct phases (initialization, beaconing, site
survey, bidirectional FHSS) with no explicit markers.  TLF detects
boundaries from three signals: (1)~the ratio of FIFO to register-only
transactions, (2)~the frequency-change rate in \reg{RegFrf} writes, and
(3)~RSSI reads without FIFO activity.  Sustained RSSI-only polling with
sequential frequency changes is classified as a site survey; a shift from
TX-only activity on a small channel set to interleaved TX/RX on a larger
one marks the move from beaconing to operational FHSS\@.

\paragraph{Scale}
The case-study capture contains 2.3~million SPI transactions.  Even a
filtered register dump produces over 300\,000 \reg{RegFrf} writes.  TLF
processes the full capture in seconds on commodity hardware.

\subsection{Detection of FIFO TX/RX Switches: PDU and Directionality Recovery}
\label{sec:fifo-detection}

In packet mode the FIFO mediates all payload data between MCU and front-end.
It is accessed at a single fixed SPI address (\texttt{0x00} on both
families) and is one byte wide, a shift register serializing each byte to or
from the modulator at the programmed bit rate.

\paragraph{FIFO operation}
On TX the MCU writes payload bytes (length field first in variable-length
mode) into the FIFO in Standby/Sleep, then enters Tx; the packet handler
prepends preamble and sync word, applies optional CRC, whitening and AES,
and shifts the frame to the modulator.  On RX it strips preamble and sync,
optionally decrypts, and deposits the payload back into the FIFO, signaling
the MCU via a DIO interrupt (\emph{PayloadReady} or \emph{FifoLevel}).

\paragraph{Distinguishing TX from RX on the bus}
The two directions are readily separable.  A FIFO write burst (wnr~=~1 at
\texttt{0x00}, data on MOSI) followed by a transition to Tx indicates an
outgoing payload; a FIFO read burst (wnr~=~0, data on MISO) after a
\emph{PayloadReady} interrupt or an Rx$\to$Standby transition indicates a
received one.  Automatic FIFO clearing on certain mode transitions
(e.g., Rx$\to$Tx) gives further framing cues.

\subsection{Register Control and Interception}
\label{sec:register-interception}

Every SPI transaction targeting a non-FIFO address is a register access.
The decoder maps each address to its functional meaning and maintains a
shadow copy of the configuration state, updated on every intercepted
write.  This shadow state---the transceiver's intended configuration at
each instant---is the decoder's primary output.

\Cref{tab:key-registers} lists the registers of highest relevance to a
decoder, grouped by functional category.  The SX1276 FSK/OOK and SX1233
maps are closely related but \emph{not} identical, and a decoder assuming
otherwise will mis-attribute parameters: a constant $+1$ offset applies on
the SX1233 for the bit-rate, deviation and frequency triplets; some
registers sit at genuinely different addresses (notably \reg{RegRxBw});
and some exist on only one part (\reg{RegDataModul} and the AES bank on
the SX1233).  The SX1276 LoRa modem further overlays different semantics
onto the same addresses when \emph{LongRangeMode} is set, so each access
must be interpreted against the active overlay (\cref{sec:multi-family}).

\begin{table}[t]
  \centering
  \caption{Key registers for bus-level parameter extraction
           (SX1276 FSK/OOK and SX1233).}
  \label{tab:key-registers}
  \resizebox{\columnwidth}{!}{%
  \scriptsize
  \setlength{\tabcolsep}{2pt}
  \begin{tabular}{@{}lll@{}}
    \toprule
    \textbf{Category} & \textbf{Registers (addr)} & \textbf{Extracted parameter} \\
    \midrule
    Op.\ mode
      & \reg{RegOpMode} (\texttt{0x01})
      & Sleep/Stdby/FS/Tx/Rx \\
    Modulation
      & \reg{RegDataModul} (\texttt{0x02})$^\dagger$
      & FSK/OOK, shaping, cont./pkt \\
    Bit rate
      & \texttt{RegBitrate\{Msb,Lsb\}} (\texttt{0x02--03})$^\ddagger$
      & $\mathit{BR}\!=\!F_\mathrm{XOSC}/\mathit{BitRate}(15{:}0)$ \\
    Freq.\ dev.
      & \texttt{RegFdev\{Msb,Lsb\}} (\texttt{0x04--05})$^\ddagger$
      & $F_\mathrm{DEV}\!=\!F_\mathrm{STEP}\!\times\!\mathit{Fdev}(13{:}0)$ \\
    Carrier freq.
      & \texttt{RegFrf\{Msb,Mid,Lsb\}} (\texttt{0x06--08})$^\ddagger$
      & $F_\mathrm{RF}$ per \cref{eq:frf} \\
    Preamble
      & \texttt{RegPreamble\{Msb,Lsb\}}
      & Preamble len.\ (bytes/sym.) \\
    Sync word
      & \reg{RegSyncConfig}, \texttt{Value1--8}
      & Network ID, length \\
    Pkt.\ format
      & \reg{RegPacketConfig1--2}
      & Fixed/var., CRC, whitening \\
    Payload len.
      & \reg{RegPayloadLength}
      & Expected/max payload size \\
    PA \& power
      & \reg{RegPaConfig} (\texttt{0x09})\,/\,\reg{RegPaLevel} (\texttt{0x11})$^\S$
      & Output power, PA select. \\
    Rx BW
      & \reg{RegRxBw} (\texttt{0x12}\,/\,\texttt{0x19})$^\S$
      & Channel filter BW \\
    AES key
      & \reg{RegAesKey1--16} (\texttt{0x3E--4D})$^\dagger$
      & 128-bit encryption key \\
    \bottomrule
    \multicolumn{3}{@{}p{1.25\columnwidth}@{}}{\scriptsize
      $\dagger$\,SX1233 only.  On SX127x \emph{DataMode} lives in
      \reg{RegPacketConfig2} (\texttt{0x31}) and there is no AES bank. \quad
      $\ddagger$\,Addr.\ offset $+1$ on SX1233. \quad
      $\S$\,SX127x\,/\,SX1233, respectively.} \\
  \end{tabular}}%
\end{table}

\paragraph{Tracking configuration state over time}
The decoder maintains a time-stamped shadow register file so that each
FIFO transaction is associated with the configuration active at that
moment.  Detecting \emph{changes} matters equally: a mid-session
\reg{RegFrf} write signals a frequency hop
(\cref{sec:sync-freq-detection}), a \reg{RegOpMode} write with a new
\emph{Mode} field marks a Tx/Rx transition (\cref{sec:fifo-detection}),
and a \reg{RegSyncValue} write during operation may indicate syncword
rotation---a deliberate anti-interception measure.  Diffing successive
snapshots yields a concise event log of exactly what changed and when.

\subsection{Frame Rebuilding from Bus Captures}
\label{sec:frame-rebuilding}

The FIFO sits \emph{inside} the radio-level framing, so bus-visible bytes
are never the over-the-air frame.  On transmit they are the payload as the
MCU supplies it, \emph{before} the packet handler prepends preamble and
sync word, appends CRC, and applies whitening, Manchester coding or AES;
on receive they are what \emph{remains after} the handler has stripped
those fields.  Reconstructing the air frame means re-applying that
transform in the appropriate direction from the preamble, sync word,
\emph{CrcOn} and \emph{DcFree} settings in the shadow state---reproducing
any enabled whitening, coding or cipher rather than assuming it absent.
TLF concatenates multi-burst fragments within one mode-transition bracket
and emits a \texttt{RecoveredPacket} bound to an immutable
\texttt{ConfigSnapshot}: payload plus the RF parameters under which it
crossed the air.

\subsection{Detection of Sync and Frequency Settings}
\label{sec:sync-freq-detection}

Three classes of configuration are especially valuable to a decoder because
they directly constrain the over-the-air waveform and, together, are
sufficient to tune a matched receiver: the sync word, preamble length, and
carrier frequency.

\paragraph{Sync word and preamble extraction}
In FSK/OOK the variable-length sync word (1--8~bytes) lives in
\reg{RegSyncValue1}--\texttt{8} (active length: \emph{SyncSize} in
\reg{RegSyncConfig}; recognition gated by \emph{SyncOn}); LoRa places a
single byte at \reg{RegSyncWord} (\texttt{0x39}), \texttt{0x34} being
reserved for LoRaWAN public networks, so its presence strongly indicates
use of that convention.  Preamble length is a 16-bit value in
\texttt{RegPreamble\{Msb,Lsb\}} (LoRa adds 4.25 symbols).  Writes to these
registers expose the network identifier and time-on-air parameters.

\paragraph{Carrier frequency and hop detection}
The carrier frequency is programmed via a 24-bit register triplet
(\reg{RegFrfMsb}, \reg{RegFrfMid}, \reg{RegFrfLsb}) and is
computed as:
\begin{equation}\label{eq:frf}
  F_\mathrm{RF} = F_\mathrm{STEP} \times \mathit{Frf}(23{:}0),
  \qquad
  F_\mathrm{STEP} = \frac{F_\mathrm{XOSC}}{2^{19}}
\end{equation}
With the standard 32\,MHz crystal this yields a resolution of
approximately 61\,Hz per step.  A decoder that tracks every write to
the \reg{RegFrf} triplet therefore reconstructs the complete frequency
history of the session.

Frequency hopping manifests differently depending on whether the
transceiver's built-in FHSS engine is used or the firmware implements its
own schedule:

\begin{itemize}
  \item \textbf{LoRa built-in FHSS.}
    The SX1276 LoRa modem provides a hardware-assisted hop mechanism
    controlled by \emph{FreqHoppingPeriod} (\reg{RegHopPeriod},
    \texttt{0x24}).  When non-zero the modem asserts the
    \emph{FhssChangeChannel} interrupt every \emph{FreqHoppingPeriod}
    symbols, and the current hop index is readable in
    \emph{FhssPresentChannel} (\reg{RegHopChannel[5:0]},
    \texttt{0x1C}).  In this mode the firmware must service each
    interrupt by writing new \reg{RegFrf} values, so every hop
    produces a cluster of three register writes immediately following
    the interrupt---a distinctive, periodic pattern on the bus that a
    decoder can lock onto.

  \item \textbf{Firmware-driven (custom) FHSS.}
    On transceivers lacking a built-in hop engine (e.g., SX1233)
    or when the firmware chooses to manage hopping in software, the MCU
    executes the hop by transitioning the transceiver through an
    intermediate mode (Rx$\to$Standby or Rx$\to$FS), reprogramming
    \reg{RegFrf}, and re-entering Tx or Rx.  The SX1233 datasheet
    prescribes exactly this sequence for spectral-purity reasons.
    Because these hops are asynchronous to any transceiver-side timer,
    the inter-hop interval must be measured from the bus timestamps
    rather than inferred from a register setting.  The decoder
    collects a time-stamped list of all \reg{RegFrf} writes and
    derives the channel table, hop period, and dwell time from their
    statistics.
\end{itemize}

In both cases the decoder's output is a channel table (set of
$F_\mathrm{RF}$ values), a hop sequence (order of visits), and
timing statistics (mean and jitter of inter-hop intervals).  These
are sufficient to replay an observed schedule, configure a receiver over
the recovered channel set, or support a real-time follower while hop
writes remain observable.

\subsection{Modeling Parameters as Generalized Objects}
\label{sec:parameter-modeling}

Family-specific decoders emit into a common, transceiver-agnostic
intermediate representation so that downstream tools (diffing,
serialization, GNU~Radio flowgraph generation, receiver-configuration
scripts) need not depend on register-level details.  The model covers
sync word, preamble, modulation (type/shaping/BW), band plan (frequencies,
hop sequence, timing, per-channel parameter variants such as rotating sync
words), encryption (key, mode, enable), whitening (polynomial, initial
state), and FEC (rate, block size, scheme).

\subsection{Supporting Multiple Transceiver Families}
\label{sec:multi-family}

Adding a family means supplying a register map, power-on-reset defaults,
and the family's mode and FIFO semantics.  The decoder self-registers under
a family name and every frontend resolves it by name without further
change.  TLF ships decoders for the SX1233 and SX127x families.  Coverage
matters because the population is heterogeneous: the same component
database also places Semtech/Xemics XE1205 transceivers in Lancet and Zala
UAV subsystems~\cite{DIU2024XE1205Lancet}---a narrowband packet-radio part,
not a LoRa device, and thus a distinct decoder target.

The SX127x is the more demanding, and instructive about what the
abstraction must absorb.  It carries \emph{two} register overlays in one
address space, selected by \emph{LongRangeMode} (\reg{RegOpMode} bit~7).
In LoRa mode the same addresses mean something else entirely:
\texttt{0x1D}--\texttt{0x1E} become modem-configuration registers encoding
spreading factor, bandwidth and coding rate, and \texttt{0x39} the
single-byte LoRa sync word.  A decoder that fixes one interpretation at
load time is therefore wrong for half of any capture that switches modes.
TLF instead resolves each access against the overlay implied by the
current shadow value of \reg{RegOpMode}.  Framing differs too: the SX1233
FSK path streams bytes through a byte-wide 66-byte FIFO and must buffer
across mode transitions, whereas the LoRa modem stages packets in its
256-byte buffer (\cref{sec:challenges}).

\subsection{Application-Layer Protocol Decoding}
\label{sec:protocol-layer}

Chip-level output is a recovered payload plus the RF configuration under
which it crossed the air.  Where the framing above that boundary is a
documented protocol, the payload need not stay opaque.  TLF therefore adds
a second plugin layer: a protocol decoder implements a \texttt{detect}
heuristic and a \texttt{decode} that yields a structured PDU, and
registers itself by name.  The layer is \emph{chip-agnostic}---it consumes
the same intermediate packet objects whichever family decoder produced
them---so a new protocol becomes available to every chip decoder at once,
and a new chip family inherits every protocol decoder.  Absent an operator
choice, each registered detector is offered the recovered packets and the
first to accept is used.

The layering also makes the trust boundary explicit: chip-level recovery is
unconditional, whereas protocol decoding may additionally need key
material---present on the bus precisely when the transceiver's own crypto
engine is used (\cref{sec:encryption}), absent when the protocol encrypts
in firmware.  In the latter case addressing, routing metadata and framing
are still recovered.

\section{Case Study Evaluation}\label{sec:evaluation}

\subsection{Target and Setup}\label{sec:cs1-setup}

The target is an SX1233-based modem forming part of a UAV C2 link,
operating in the US~915\,MHz ISM band.  The system employs read-out
protection, preventing firmware extraction without family-specific
hardware attacks (e.g., glitching~\cite{9505146} or decapping).  A DSLogic logic analyzer
captured approximately 110\,s of SPI traffic (20\,MHz sample rate,
2.3~million transactions), spanning initialization through steady-state
bidirectional operation.

\subsection{Recovered Parameters}\label{sec:cs1}

\begin{table}[t]
  \centering
  \caption{Extracted parameters for the SX1233 target.}
  \label{tab:fhss-case1}
  \footnotesize
  \setlength{\tabcolsep}{4pt}
  \begin{tabular}{@{}lll@{}}
    \toprule
    \textbf{Category}  & \textbf{Parameter}       & \textbf{Value} \\
    \midrule
    \multirow{5}{*}{RF config}
      & Modulation               & GFSK (BT\,=\,1.0) \\
      & Bit rate                 & 76\,190\,bps \\
      & Freq.\ deviation         & 80\,kHz \\
      & Rx bandwidth             & 200\,kHz \\
      & TX power                 & +10\,dBm \\
    \midrule
    \multirow{4}{*}{Packet}
      & Preamble                 & 3\,bytes \\
      & Sync word length         & 4\,bytes \\
      & Payload size             & 113\,B (beacon) / 153--158\,B (op.) \\
      & HW encryption            & None (AesOn\,=\,0) \\
    \midrule
    \multirow{5}{*}{Hopping}
      & Operational band         & 912.0--917.9\,MHz \\
      & Channel count            & 55 (union); 44 per direction \\
      & Channel spacing          & 100\,kHz \\
      & Beacon hop interval      & 31.1\,ms (3 channels) \\
      & Hop sequence             & Aperiodic over capture window \\
    \midrule
    \multirow{4}{*}{\shortstack[l]{Sync word\\rotation}}
      & Unique sync words        & 167 (166 rotating $+$ 1 beacon) \\
      & Burst windows            & 2 (at ${\sim}$47\,s and ${\sim}$82\,s) \\
      & Freq.\ binding           & 1:1 \\
      & Reuse                    & Each pair used $\leq$\,2$\times$ \\
    \midrule
    \multirow{3}{*}{Site survey}
      & Survey band              & 894.0--926.0\,MHz \\
      & Channels surveyed        & 128 \\
      & Measurements / channel   & $\sim$\,2\,417 (mean) \\
    \bottomrule
  \end{tabular}
\end{table}

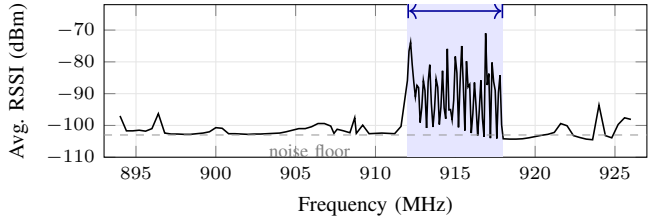
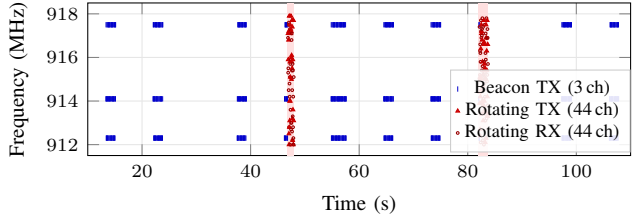
\begin{figure}[t]
  \centering
  \begin{subfigure}[t]{\columnwidth}
    \centering
    \begin{tikzpicture}
      \begin{axis}[
        width=\columnwidth,
        height=3.6cm,
        xlabel={Frequency (MHz)},
        ylabel={Avg.\ RSSI (dBm)},
        xmin=893, xmax=927,
        ymin=-110, ymax=-62,
        ytick={-110,-100,-90,-80,-70},
        grid=major,
        grid style={gray!20},
        tick label style={font=\scriptsize},
        label style={font=\footnotesize},
      ]
        \fill[blue!10] (axis cs:912,-110) rectangle (axis cs:918,-62);
        \addplot[mark=none, semithick, color=black]
          table[x=freq_mhz, y=rssi_avg_dbm, comment chars=\%]
          {figs/rssi_survey.dat};
        \draw[dashed, black!40, thin]
          (axis cs:893,-103) -- (axis cs:927,-103);
        \node[font=\scriptsize, black!50, anchor=north east]
          at (axis cs:909,-103) {noise floor};
        \draw[semithick, blue!60!black, |<->|]
          (axis cs:912,-64) -- (axis cs:918,-64);
        \node[font=\scriptsize, blue!60!black, anchor=south]
          at (axis cs:915,-64.5) {operational band};
      \end{axis}
    \end{tikzpicture}
    \caption{RSSI site survey (309\,329 reads over 128 channels,
             ${\sim}$2\,417 per channel).  The
             912--918\,MHz operational band (shaded) shows elevated RSSI
             from co-located transmitter activity.}
    \label{fig:rssi-survey}
  \end{subfigure}

  \vspace{4pt}

  \begin{subfigure}[t]{\columnwidth}
    \centering
    \begin{tikzpicture}
      \begin{axis}[
        width=\columnwidth,
        height=3.6cm,
        xlabel={Time (s)},
        ylabel={Frequency (MHz)},
        xmin=10, xmax=110,
        ymin=911.5, ymax=918.5,
        grid=major,
        grid style={gray!20},
        tick label style={font=\scriptsize},
        label style={font=\footnotesize},
        scatter/classes={
          0={mark=|, mark size=1.0pt, line width=0.3pt, blue!80!black},
          1={mark=triangle*, mark size=0.8pt, red!80!black},
          2={mark=o, mark size=0.5pt, draw=red!60!black, thin}
        },
        legend style={
          at={(1.00,0.02)}, anchor=south east,
          font=\scriptsize,
          draw=gray!40, thin,
          fill=white, fill opacity=0.9, text opacity=1,
          row sep=-2pt, inner sep=2pt,
        },
      ]
        \fill[red!12] (axis cs:46.8,911.5) rectangle (axis cs:48.0,918.5);
        \fill[red!12] (axis cs:82.0,911.5) rectangle (axis cs:83.8,918.5);
        \addplot[scatter, only marks, scatter src=explicit]
          table[x=time_s, y=freq_mhz, meta=phase, comment chars=\%]
          {figs/hop_timeline.dat};
        \legend{Beacon TX (3\,ch),
                Rotating TX (44\,ch),
                Rotating RX (44\,ch)}
      \end{axis}
    \end{tikzpicture}
    \caption{Frequency hopping timeline.  Beacons (blue) hop across 3
             channels at 31\,ms cadence.  Two burst windows (shaded)
             interleave RX/TX pairs with per-packet sync word rotation.
             Each direction visits 44 channels; their union is the full
             55-channel table.}
    \label{fig:hop-timeline}
  \end{subfigure}

  \caption{FHSS characterization of the SX1233 target, extracted entirely
           from SPI bus traffic.}
  \label{fig:fhss-case1}
\end{figure}

\Cref{tab:fhss-case1} summarizes the RF configuration and FHSS
characteristics extracted from the initialization and operational
sequences.  No use of the SX1233's built-in AES engine was detected
(\emph{AesOn}\,=\,0), establishing that payloads are not
hardware-encrypted.  This is consistent with coding and confidentiality
being implemented in firmware above the transceiver layer, though the bus
alone cannot establish which of FEC, whitening or encryption are present.
\Cref{fig:fhss-case1} visualizes the site-survey spectrum and frequency
hopping timeline. In both cases, TLF made the insights or discoverables
\emph{immediately obvious without any operator intervention}.

\paragraph{Beaconing phase}
The initial phase is beacon transmission: 113-byte packets at a fixed
${\sim}31$\,ms interval cycling across three frequencies (912.3, 914.1,
917.5\,MHz) with a constant sync word.  The decoder identifies it from the
uniform packet size, fixed sync word, and absence of Rx-mode transitions.

\paragraph{Site survey}
Prior to entering the operational phase the firmware performs an RF
environment survey.  The decoder detected 309\,329 RSSI register reads
across 128 frequencies spanning 894--926\,MHz, a mean of ${\sim}2\,417$
measurements per channel.  This behavior---sequential frequency tuning,
RSSI polling, no FIFO activity---is automatically flagged by the framework
as a site survey, and the resulting noise-floor profile is included in the
output.  Programmatic surveying suggests environment-aware design: the
modem profiles the band before operating, which could support runtime
channel avoidance.

\paragraph{Operational FHSS phase}
After the survey the modem transitions to bidirectional communication with
frequency hopping across 55 channels (912.0--917.9\,MHz, 100\,kHz
spacing).  The decoder recovered 682 packets (both TX and RX), extracted
the full channel table, and computed hop timing statistics: a mean dwell
time of ${\sim}139$\,ms with a minimum of ${\sim}14$\,ms.

Two features distinguish this phase from a conventional FHSS link: the
sync word changes on every packet (each rotating packet uses a distinct
observed sync word, implying state that evolves in lockstep between
peers), and the
payloads are opaque, consistent with the absence of built-in AES usage.
The capture does not, however, identify the \emph{generator} of that
state: a keystream, a counter, a seeded PRNG and a precomputed table are
all equally consistent with the observations, and distinguishing them
would require the firmware analysis this method avoids.  We therefore
describe the rotation as peer-synchronized rather than cipher-derived.

TLF's protocol classifier automatically groups the 682~packets into three
classes by direction, sync word behavior, and payload size: fixed-syncword
TX beacons (516~packets, 113\,B, 3~channels), rotating-syncword TX
(82~packets, 158\,B, 44~channels), and rotating-syncword RX (84~packets,
153--154\,B, 44~channels).  Turnaround times of 14.4\,ms RX$\to$TX and
17.5\,ms TX$\to$RX confirm the half-duplex structure, and beacon payloads
share a 15-byte common prefix---likely a fixed header with a counter or
timestamp.  Two counts warrant explanation: rotating TX visits 44 channels
and rotating RX 44, but the sets differ and their \emph{union} is exactly
the channel table, so the full plan is recoverable only by observing
both directions; and the 166 rotating packets carry 166 distinct sync
words, one apiece, the fixed beacon value bringing the total to 167.

\subsection{Case Study II: SX1276 LoRa and Application-Layer Decode}
\label{sec:cs2}

The second target is a Meshtastic~\cite{MeshtasticOffGrid} node built around an
SX1276, captured with the same equipment and workflow.  It exercises a
second chip family, the LoRa overlay, and the protocol layer of
\cref{sec:protocol-layer}.  It is also a deliberately \emph{favorable}
case---the protocol is open and its default channel key published---which
is what makes it useful: it isolates what bus-level recovery contributes
independently of cryptanalysis.

\paragraph{Chip-level recovery}
From a cold-boot capture of 117 SPI transactions (100\,MHz sampling,
decoded in 4.0\,s) TLF recovered the full LoRa modem configuration:
906.875\,MHz carrier, SF11, 250\,kHz bandwidth, coding rate 4/5,
single-byte sync word \texttt{0x2B}, 16-symbol preamble, PA 15\,dBm,
resting in \emph{RxContinuous}.  A longer cold-boot capture
($\sim$50\,s, 1\,507 transactions, 17.6\,s) reproduced it independently.
The LoRa-specific subset---spreading factor, bandwidth, coding rate,
preamble, and sync word---would be misinterpreted as FSK/OOK state by an
overlay-blind decoder; carrier frequency, PA configuration and
\reg{RegOpMode} are common to both overlays.

The two cold-boot captures agree with one another, and a configuration read
out independently through the node's management interface matched the
recovered modem parameters.  The recovered sync word \texttt{0x2B} likewise
matches the value Meshtastic documents, and successful parsing and
decryption of every recovered packet validate the configuration in use.

\paragraph{Configuration recovery requires a cold boot}
Firmware programs the modem once, at initialization, and thereafter
touches only the FIFO and a few status and mode registers.  Captures
starting later---including a $\sim$3\,min trace of the same node---contain
\emph{zero} \reg{RegFrf} writes, leaving those shadow registers at
power-on-reset defaults.  TLF reports the register-write count so this is
visible rather than silent: a configuration resting at POR values with no
observed writes is an artifact of capture timing, not a measurement.  An
intercept must therefore span a power cycle---for a captured platform, the
analyst controls when it boots.  On this target, later captures still
supported FIFO packet extraction and Meshtastic parsing once the chip
family and LoRa mode were known, although complete configuration
attribution remained unavailable.

\paragraph{Application-layer decode}
On the 50-second capture the detector selected the Meshtastic decoder
without operator input and all 27~recovered packets parsed as valid PDUs.
Headers---source and destination, packet ID, hop limit and start, channel
hash---decode unconditionally, being sent in the clear.  Of the 27,
21~payloads additionally decrypted under the published default channel
key, yielding node-identity announcements, position reports, range-test
sequences and a traceroute exchange across three distinct nodes.  The
remaining 6 did not decrypt and are reported as such rather than guessed
(direct messages under Meshtastic's X25519 scheme, for which no key is on
the bus).

That split is the honest shape of the result: not a break of Meshtastic,
since the decrypted traffic used a key public by design.  It illustrates
the layering of \cref{sec:protocol-layer}---chip recovery is
unconditional, header decoding follows from the specification, and payload
decryption succeeds exactly when key material is independently available.

\section{Related Work}\label{sec:related-work}

To our knowledge, prior work has not systematized MCU--transceiver bus
interception as a stateful method for recovering RF configuration and
runtime transceiver behavior.  Existing approaches fall into three
categories.

\paragraph{Firmware RE}
Binary analysis (e.g., Ghidra~\cite{ghidra2019}) requires firmware
extraction via JTAG/SWD or glitching~\cite{9505146}, both
time-consuming and target-specific.  Emulation
approaches~\cite{pucher2020avrs,EURECOM-5437} face the same hurdle on
proprietary architectures.  Bus-protocol detection
work~\cite{11391743} reduces some of this effort but still requires
firmware in hand.  TLF sidesteps extraction entirely.

\paragraph{Signal-level / SDR}
GNU~Radio~\cite{gnuradio} and Universal Radio
Hacker~\cite{pohl2018universal} recover modulation, symbol rate, and
sometimes packet structure from the air, but require proximity, tolerate
channel conditions poorly, and---for FHSS---must track hops before any
channel can be decoded.

\paragraph{Hardware / bus-level}
Commercial (Saleae~\cite{saleae}) and open-source
(sigrok~\cite{sigrok}) protocol decoders provide raw byte-level decoding
without transceiver semantics.  Prior bus-level attack work has focused on
firmware extraction~\cite{costin2014large} or side channels.  TLF adds
the stateful semantic layer that maps raw transactions onto the
transceiver's register model.  Complementary LoRa security and
link-denial work~\cite{aras2017exploring,Wu2026PreambleInjectionJamming}
assumes demodulated packets or a known waveform.  TLF supplies
those parameters from register writes.

\section{Discussion and Limitations}\label{sec:discussion}

The approach has four intrinsic limitations: it requires physical access or
privileged emulation of target hardware; firmware-level behavior (custom
FEC, whitening, encryption, or the logic driving sync word rotation and
FHSS schedules) is observable at the transceiver boundary but not
explainable from the bus alone; each transceiver family requires an
explicit decoder (a task comparable in scope to writing a sigrok protocol
decoder); and configuration recovery depends on capture timing, since
firmware that programs the modem only at initialization leaves a
mid-session trace with no configuration writes to observe
(\cref{sec:cs2})---recovered configuration must therefore always be read
alongside the count of observed writes, which TLF reports.  Packet and
protocol recovery survived on this target once the chip family and mode
were known.  None is intractable: captured platforms routinely supply
physical access and need not be functional---salvageable control circuitry
suffices---and the hardware (a sub-\$300 logic analyzer and a laptop) is
portable enough for forward deployment.  Against firmware-extraction-based
characterization, which takes days or weeks per target and demands
architecture-specific expertise, TLF produces useful parameters in minutes
from a single capture.

\section{Conclusion}\label{sec:conclusion}

TLF shows that a stateful decoder built around a shadow register file can
transform millions of raw SPI transactions into a structured RF-domain
representation of transceiver behavior, without firmware extraction or
signal processing.  Attributing every FIFO transaction with the exact RF
configuration active at that instant is what enables automatic phase
detection, FHSS analysis, and sync word rotation characterization at scale.
On the SX1233 case study, the full register-exposed
configuration---band plan, hop timing, beaconing protocol,
site-survey behavior---was recovered from a single 110-second capture
processed in under 60\,s.  On an SX1276 Meshtastic node the same framework
resolved the LoRa overlay, recovered the modem configuration from a
cold-boot trace, and drove a pluggable protocol stage that parsed every
recovered packet as an application PDU\@.  In both cases firmware-level
cryptographic state stayed opaque, as expected: neither system asks its
transceiver to do cryptography, so no key ever crosses the bus.
Bus-level interception recovers what the transceiver is told to do, not
what the firmware decided---a boundary that is architectural rather than
incidental.  TLF includes preliminary support for GNU~Radio
flowgraph generation and IQ synthesis.  Future work extends decoders, adds
support for a shadow command-parameter model (e.g., for the SX126x family),
adds protocol plugins beyond Meshtastic and multi-capture correlation
across paired devices, and targets a real-time field tool.

\makeatletter
\g@addto@macro\UrlBreaks{%
  \do\a\do\b\do\c\do\d\do\e\do\f\do\g\do\h\do\i\do\j\do\k\do\l\do\m%
  \do\n\do\o\do\p\do\q\do\r\do\s\do\t\do\u\do\v\do\w\do\x\do\y\do\z%
  \do\0\do\1\do\2\do\3\do\4\do\5\do\6\do\7\do\8\do\9}
\makeatother
\Urlmuskip=0mu plus 1mu\relax

\bibliographystyle{IEEEtran}
\bibliography{transceiver-config-lifting}

\end{document}